\documentclass[prb, superscriptaddress, twocolumn]{revtex4}
\usepackage{color, graphicx, amssymb, amsmath}  

\begin{document}

\author{Nicholas C. Koshnick}  
\affiliation{Departments of Physics and Applied Physics, Stanford  
University, Stanford, CA 94305}  
\author{Martin E. Huber}  
\affiliation{Departments of Physics and Electrical Engineering, University of Colorado Denver,   
Denver, CO 80217}  
\author{Julie A. Bert}  
\affiliation{Departments of Physics and Applied Physics, Stanford  
University, Stanford, CA 94305}  
\author{Clifford W. Hicks}
\affiliation{Departments of Physics and Applied Physics, Stanford  
University, Stanford, CA 94305}  
\author{Jeff Large}
\affiliation{Circuit Design Repair Lab, Texas Instruments, Dallas, TX 75243}
\author{Hal Edwards}
\affiliation{Circuit Design Repair Lab, Texas Instruments, Dallas, TX 75243}
\author{Kathryn A. Moler}  
\email{kmoler@stanford.edu}  
\affiliation{Departments of Physics and Applied Physics, Stanford  
University, Stanford, CA 94305}  

\title{A Terraced Scanning Superconducting Quantum Interference Device Susceptometer with Sub-Micron Pickup Loops}

\begin{abstract}
Superconducting Quantum Interference Devices (SQUIDs) can have excellent spin sensitivity depending on their magnetic flux noise, pick-up loop diameter, and distance from the sample.  We report a family of scanning SQUID susceptometers with terraced tips that position the pick-up loops 300 nm from the sample.  The 600 nm -- 2$\mu$m pickup loops, defined by focused ion beam, are integrated into a 12-layer optical lithography process allowing flux-locked feedback, in situ background subtraction and optimized flux noise.  These features enable a sensitivity of $\sim$70 electron spins per root Hertz at 4K.
\end{abstract}

\maketitle
  
In 1989, Ketchen et al$.$\cite{ketchen_design_1989} argued that the advent of sub-micron lithography should enable Superconducting Quantum Interference Devices (SQUIDs) with single electron spin sensitivity.  Stationary devices can measure nanomagnets with great success \cite{wernsdorfer_experimental_1997}, but it remains difficult, even with the highest spin-sensitivity SQUIDs \cite{cleuziou_carbon_2006}, to detect single-molecule field sources.  Scanning devices  \cite{vu_design_1993, kirtley_high-resolution_1995, morooka_development_2000,hasselbach_microsquid_2000, freitag_electrically_2006, huber_gradiometric_2008} make it possible to isolate magnetic field sources and to perform background measurements in situ \cite{koshnick_fluctuation_2007}. Single-layer scanning nano-SQUIDs \cite{hasselbach_microsquid_2000} have not yet achieved the excellent flux sensitivity of the best large multi-layer SQUIDs. Multilayer SQUIDs thus far have had large pickup areas that do not capture dipole field lines well thereby limiting the spin sensitivity of these devices.  For a normal-oriented dipole on the center line of a pickup loop of radius $R$, the spin sensitivity, $S_n$, is 
\begin{equation}
S_n (\mu_B/\sqrt{\textrm{Hz}}) = \Phi_n \frac{R}{r_e} \left( 1 + \frac{h^2}{R^2} \right)^{3/2}
\label{equ:spin}
\end{equation}
where $\Phi_n$ is the flux noise in units of $\Phi_0/\sqrt{\textrm{Hz}}$, $h$ is the pickup loop's height above the sample, and $r_e \approx 2.82 \times 10^{-15}$ m \cite{ketchen_design_1989}. On the center line, near-optimal signal is achieved for $h<R$. The spin sensitivity can be further improved by placing the dipole near the edge of the pickup loop, although demagnetization limits this enhancement for $h<w$, where $w$ is the linewidth. For simplicity, we use Eq. 1 to compare representative published scanning SQUIDs (Table 1). 

\begin{table}[ht]
\caption{Survey of reported scanning SQUIDs and estimated spin sensitivity for $h = 0$ (Eq. \ref{equ:spin}).
With one exception \cite{hasselbach_microsquid_2000}, the corners  are typically 20-60 $\mu m$ from the pickup loop, likely limiting $h$ to 1 -- 3 $\mu m$. For rectangular loops we use $R = (l_1l_2/\pi)^{1/2}$. }
\centering
\begin{tabular}{  l  r  r  r  r  r  }
Principal  								& Year  	& Size	&Flux Noise& Spin Sensitivity  \\
Investigator 				&  & ($\mu$m$^2$) & ($\mu\Phi_0/\sqrt{\textrm{Hz}}$) &($\mu_B/\sqrt{\textrm{Hz}}$)  \\
\hline 
Vu \cite{vu_design_1993}		& 1993	& 100	& 3		& 6,000 \\
Kirtley \cite{kirtley_high-resolution_1995} 	&1995 	& 81 	& 2 		& 3,700  \\
Morooka \cite{morooka_development_2000} 	& 2000	& 16\cite{midp} 	& 8		& 6,400\cite{midp} \\
Hasselbach \cite{hasselbach_microsquid_2000} & 2000	& 4		& 100 	& 40,000 \\	
Freitag \cite{freitag_electrically_2006} 	& 2006 	&12 		& 2 		& 1,400	 \\		
Huber \cite{huber_gradiometric_2008} 		& 2008	& 12  	& 0.8\cite{4knoise}& 640\cite{4knoise}\\
Present Work			 				& 2008	& 0.3  	& 0.7 	& 74  \\

\end{tabular}
\label{spintable}
\end{table}

Our scanning SQUID combines Focused Ion Beam (FIB) defined pick-up loops with a 12 layer optical lithography process that includes local field coils. Integrated terraces minimize $h$. We characterize the imaging kernel with a superconducting vortex and a dipole field source. Flux noise measurements at 4 Kelvin demonstrate a spin sensitivity of $\sim$70 $\mu_B/\sqrt{\textrm{Hz}}$. Flux noise may decrease at lower temperatures \cite{huber_gradiometric_2008, SOM} leading to a projected sensitivity of  $\lesssim$ 15 $\mu_B/\sqrt{\textrm{Hz}}$.

Our susceptometer incorporates two symmetric counter-wound arms, each with an integrated modulation loop, pickup loop, and local field coil (Fig 1a). A three metalization layer, linear coaxial transmission line geometry shields the device from magnetic fields. The transmission line geometry has a low inductance per unit length ($\sim$10 pH/mm), which allows for a large separation between the feedback/junction area and the two pickup loops without significantly increasing the devices theoretical white noise floor \cite{tesche_dc_1977}. The separation permits the use of standard, well optimized junction and resistive shunt fabrication processes \cite{sauvageau_superconducting_1995}. The resistive shunts ensure a non-hysteretic response.  The scanning SQUID is voltage biased and its current is amplified with a SQUID Series Array (SSA) amplifier \cite{SAS}.  A feedback circuit controls the current in the modulation loop, responding to the SSA output voltage to create a flux locked loop.   Feedback linearizes the signal and allows for optimal sensitivity at all applied fields. The counter-wound field coils aid background subtraction \cite{huber_gradiometric_2008}. By applying a local field to the sample only in the area of the pickup loop, the field coils also allow for a low magnetic field environment near the junction, modulation, and amplification stages. 

To achieve optimal flux noise \cite{tesche_dc_1977}, each junction's critical current, I$_0$, is approximately half the superconducting flux quantum, $\Phi_0/2$, divided by the SQUID's self inductance, L.  At 4 K, we have a 0.7 $\mu \Phi_0/\sqrt{\textrm{Hz}}$ noise floor above 300 Hz and 1.2 $\mu\Phi_0/\sqrt{\textrm{Hz}}$ 1/f-like noise at 10 Hz (Fig. 2f). If the dominant flux noise is Johnson noise in the shunt resistors, as indicated by $T^{1/2}$ temperature dependence in a previous similar device \cite{huber_gradiometric_2008},  a white noise floor of 0.25 $\mu \Phi_0/\sqrt{\textrm{Hz}}$ may be achievable at 300 mK \cite{SOM}. Cooling fins attached to the shunt resistors of some devices to minimize the effect of electron-phonon coupling limited cooling may enable a white noise floor of 0.12 $\mu \Phi_0/\sqrt{\textrm{Hz}}$ at dilution refrigerator temperatures \cite{SOM}. 

When limited by Johnson noise in resistive shunts, the theoretical flux noise dependence scales \cite{tesche_dc_1977} like $L^{3/2}$, whereas quantum noise scales like $L^{1/2}$.  The incentive for a well quantified low inductance adds to the criteria for optimal pickup loop design.  When the width, $w$, or the thickness, $t$, of a superconducting feature become smaller than the penetration depth, $\lambda$, kinetic inductance can overcome the geometrical inductance contribution \cite{brandt_superconducting_2004}  and scales like $L_k \propto \lambda^2/wt$ \cite{majer_simple_2002}. Thus, linewidths smaller than $\lambda$ are undesirable. This effect, along with phase winding considerations related to coherence length effects  \cite{hasselbach_micro-superconducting_2002}, ultimately sets the pickup loop size limit.  Inductance also scales with feature length,  so we have kept the sub-micron portion of the leads short, just long enough to allow the pickup loop to touch down first without excessive stray pickup.

For optimal coupling, a dipole on the center line of the pickup loop should have $h<R$, while a dipole near the edge of the pickup loop should have $h<w$.   
Fig. 1b shows a optically defined, w = 0.6 $\mu$m,  R = 3.2 $\mu$m, pickup loop pattern with etch features inside and outside the field coil.  The outer etch supplements hand polishing to bring the corner of the chip close to the field coil, and the inner etch reduces the oxide layer above the field coil.  The thickness of the multiple layers are important parts of the design.  In Fig. 1b, the pickup loop is under 250 nm of Si0$_2$ as required for a top layer of shielding \cite{SOM}.  It is thus at least this distance from the surface.  The well created by the circular field coil allows little tolerance from the optimal alignment angle of 2.5 degrees (Inset Fig. 1b).  Additionally, it is difficult to align the device such that the off-center field coil leads don't touch down first. While the SiO$_2$ layer and limited alignment tolerance is suitable for the $w$ and $R$ of the optically patterned design, these effects are detrimental for sub-micron pickup loops.

\begin{figure}[ht]
\centering
\includegraphics[scale=0.9]{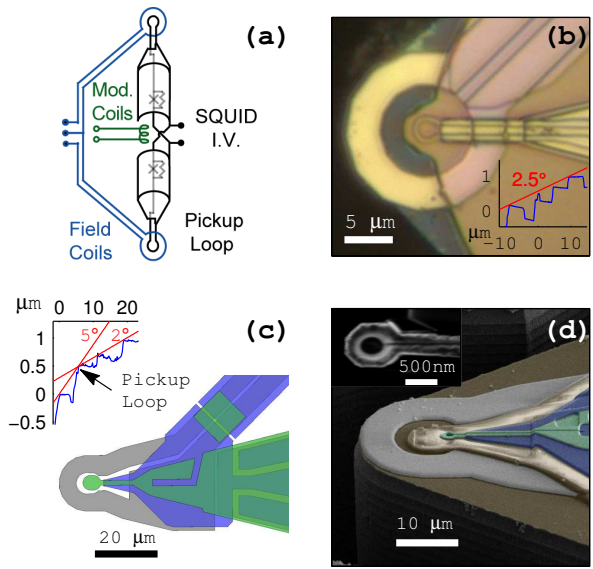}
\caption{
a) Diagram of a counterwound SQUID susceptometer.  Both the optically patterned tips (b) and FIB defined tips (c, d) feature etch defined terraces that reduce the pickup loop to sample distance.  Figure b, inset: AFM data down the center line of the device showing that the pickup loop is closest to the surface when the tip is aligned at precisely 2.5 degrees (more detail in \cite{SOM}). In the FIB design (c), the thickness of the field coil and and pickup loop leads combine with the inner terrace to form a high centerline that allows roll angle tolerance. Figure c, inset: AFM data showing the pickup loop can touch down first when the pitch angle is between 2$^\circ$-5$^\circ$.  Pickup loops down to 600 nm can be reliably fabricated with a FIB defined etch process of the topmost layer. 
}
\end{figure}

We explored several techniques to create superconducting sub-micron pickup loops integrated with the multilayer structure: ebeam defined lift-off lithography with Al, ebeam lithography for etching optically patterned Nb layers, and FIB etching of optically patterned Nb layers. The FIB etching was the most tractable. We also found that sputtered Nb has a smaller penetration depth ($\sim$90 nm) than e-beam evaporated Al patterned with PMMA liftoff ($\sim$120-160 nm), allowing for smaller linewidths and reducing the calculated  \cite{kamon_fasthenry_1994} inductance for a pair of pickup loops (22 pH vs. 66 pH). The inductance of the rest of the design is 60-65 pH. Here, we only report results from optically and FIB defined Nb tips. 
 
Our FIB design uses three superconducting layers (Fig 1c) such that the field coil lines (gray) run underneath a shielding layer (purple) and approach the tip from the same angle as the pickup loop.  The pickup loop on the top layer (green) is closest to the sample, which also allows for post-optical FIB processing. This design allows the pickup loop to touch first when the SQUID is aligned to a pitch angle of 2$^\circ$ -- 5$^\circ$(Fig. 1c inset), with a roll tolerance equal to the pitch angle. 

To increase durability, we fabricated some devices with the pickup loop retracted from the end of the etch-defined Si0$_2$ tab (Fig 1d), allowing the Si0$_2$ to take the brunt of the wear. The Si0$_2$ tab also overlaps with the inside edge of the field coil, making a high point that protects the pickup loop for pitch angles less than 2 degrees. The alignment angle is difficult to set accurately and can change due to thermal contractions, so these considerations are important for protecting the device.

\begin{figure}[ht]
\centering
\includegraphics[scale=0.9]{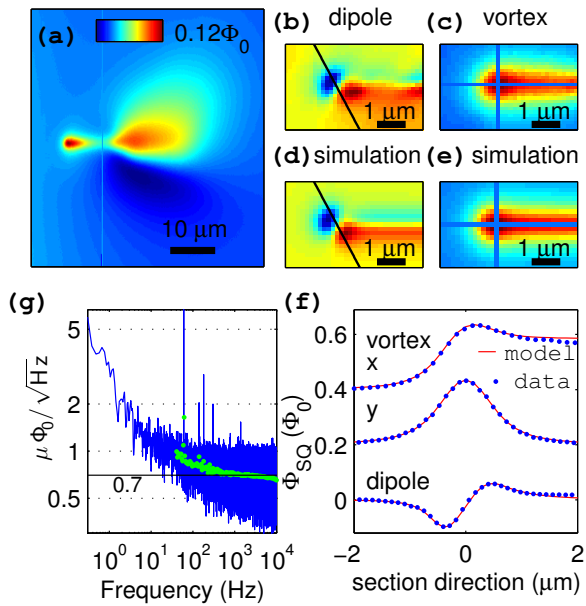}
\caption{
a) A FIB-fabricated device's magnetometry response near an isolated superconducting vortex.  Both positive and negative flux coupling occurs near the leads to the pickup loop (shown on the right side). (b-f) data and modeled results for a pickup loop with a 500 nm inner diameter and 250 nm linewidth. The flux captured from an isolated surface dipole (b) and monopole magnetic field source (c) agrees with the modeled results (d,e) calculated by integrating the field lines that thread a pickup loop kernal located 400 nm above the surface. 
(f) Linescans of (b-e) offset for clarity.  (g) Noise spectrum at 4 K.  
The green dots show averaged values and the black line displays the average between 2 KHz and 5 KHz
}
\end{figure}

We imaged Sr$_2$RuO$_4$ (Fig. 2) to characterize the FIB-defined device's coupling to a sample. Flux from a monopole-like superconducting vortex can couple through both the pickup loop and its leads (Fig. 2a). Our smallest SQUIDs are designed to do comparative studies on and off a particular mesoscopic structure, rather than provide a point like imaging kernel.

Fitting a simple model of the pickup loop response to the vortex and dipole (Fig. 2b-g) gives an effective $h$.  The vortex model is a monopole field source one penetration depth ($\lambda_{Sr_2RuO_4}$ = 150 nm) below the surface \cite{kirtley_high-resolution_1995}.   The dipole model is a free-space dipole field source at the surface. The field from each of these two sources is integrated over the effective pickup loop area at an effective height $h_{eff} = 400$ nm. This $h_{eff}$ implies that the closest side of the 200 nm thick pickup loop is 300 nm above the scanned surface.  Several effects could make this estimate of $h$ larger than the physical distance from the sample, such as the existence of a Meissner image dipole, $\lambda_{Sr_2RuO_4} >$ 150 nm due to dead layers or finite T, and demagnetization effects from the thickness of the pickup loop. 

In conclusion, we have demonstrated SQUIDs with 0.7 $\mu\Phi_0/\sqrt{\textrm{Hz}}$ flux noise at 4 K, reliable FIB pickup loops with diameters as small as 600 nm, and a terraced geometry that allows the pickup loop to come within 300 nm of the surface.   These features give a spin sensitivity of $\sim$70 $\mu_B/\sqrt{\textrm{Hz}}$, that is, the device noise is equivalent to the signal from a single electron spin after an averaging time of a little more than one hour.   At lower temperatures a lower flux noise is likely, leading to spin sensitivities less than $15 \mu_B/\sqrt{\textrm{Hz}}$.   

We acknowledge funding from NSF grants PHY-0425897, DMR-0507931, and ECS-9731293 and would like to thank Hendrik Bluhm and John Kirtley for helpful discussions.  



\end{document}